%% file: main.tex
\documentclass[runningheads]{llncs}
\usepackage[T1]{fontenc}
\usepackage{graphicx}

\usepackage{amsmath}
\usepackage{amssymb}
\usepackage{subcaption}
\usepackage[utf8]{inputenc}
\usepackage{url}
\usepackage{amsfonts}
\usepackage{xcolor}
\usepackage{ifthen}

\newcommand{\startpara}[1]{{%
\vskip6pt\noindent
{\bf #1.}}}

\usepackage{bbding}
\makeatletter
\RequirePackage[bookmarks,unicode,colorlinks=true]{hyperref}%
   \def\@citecolor{blue}%
   \def\@urlcolor{blue}%
   \def\@linkcolor{blue}%

\def\orcidID#1{\smash{\href{http://orcid.org/#1}{\protect\raisebox{-1.25pt}{\protect\includegraphics{ORCID_Color.eps}}}}}
\makeatother

\begin{document}
\title{Probabilistic Model Checking: \\ Applications and Trends}


\def\techreport

\ifthenelse{\isundefined{\techreport}}{\author{Marta Kwiatkowska\inst{1}\orcidID{0000-0001-9022-7599} \and Gethin Norman\inst{1,2}\orcidID{0000-0001-9326-4344} \and David~Parker\inst{1}(\Envelope)\orcidID{0000-0003-4137-8862}}
}{
\author{Marta Kwiatkowska\inst{1} \and Gethin Norman\inst{1,2} \and David~Parker\inst{1}}
}
\authorrunning{M. Kwiatkowska, G. Norman and D. Parker}

\institute{Department of Computer Science, University of Oxford, Oxford, UK \email{\{marta.kwiatkowska,david.parker\}@cs.ox.ac.uk}
\and School of Computing Science, University of Glasgow, Glasgow, UK
\email{gethin.norman@glasgow.ac.uk}}
 
\maketitle              
\begin{abstract}
Probabilistic model checking is an approach to the formal modelling and
analysis of stochastic systems.
Over the past twenty five years, the number of different formalisms and techniques
developed in this field has grown considerably,
as has the range of problems to which it has been applied.
In this paper, we identify the main application domains
in which probabilistic model checking
has proved valuable and discuss how these have evolved over time.
We summarise the key strands of the underlying theory and technologies
that have contributed to these advances, and highlight examples which
illustrate the benefits that probabilistic model checking can bring.
The aim is to inform potential users of these techniques
and to guide future developments in the field.
\end{abstract}

\input{intro}
\input{areas}
\input{concl}

\bibliographystyle{splncs04}
\bibliography{cbfest}

\end{document}

%% file: intro.tex
\vspace*{0.7em}

\section{Introduction}\label{sect:intro}

Probabilistic model checking~\cite{BAFK18} is a technique for the formal verification
of stochastic systems. Properties to be verified are specified in temporal logic
and then algorithmically checked against a 
model of the system.
Some of the earliest work in the field dates from the 1980s~\cite{Var85,CY88},
where algorithms were developed for computing the probability that
linear temporal logic 
specifications are satisfied by sequential or concurrent probabilistic programs.
The primary motivation was to establish the correctness of randomised algorithms,
which had proved to be difficult, particularly in the context of concurrency.

Further verification techniques for these models, which are now typically referred to as
discrete-time Markov chains (DTMCs) and Markov decision processes (MDPs),
were soon developed. The now widely used temporal logic PCTL was proposed
for DTMCs~\cite{HJ94} and MDPs~\cite{BdA95},
expanding the range of properties that could be specified.
Cited motivations included verifying the performance and reliability
of, e.g., computer networks.

An extension of PCTL to the continuous-time setting
was proposed in the late 1990s, called CSL~\cite{ASSB96}.
This was designed for continuous-time Markov chains (CTMCs),
a well established model for assessing performance and dependability properties of computing and communication systems.
Efficient model checking algorithms were also developed~\cite{BHHK03}.
The further integration of costs and rewards, well-known notions from Markov modelling,
into the probabilistic model checking framework
led to additional temporal logics~\cite{HCH+02},
allowing a modeller to reason about other quantitative characteristics such as power consumption.
The flexibility of these formalisms meant that they applied equally
well in non-traditional application domains such as biological systems.

For the model of MDPs, interest grew in synthesising ``correct-by-construction''
controllers or policies, using temporal logic to specify the desired behaviour
of the system under control.
Existing temporal logics and model checking algorithms
were extended to incorporate various cost- and reward-based measures
and multi-objective specifications~\cite{EKVY08}.
This opened up new applications, including a variety of scheduling and planning problems.
Probabilistic variants of the timed automata formalism,
extending MDPs with clocks,
allowed  modelling of stochastic systems with real-time constraints and delays~\cite{NPS13}.

Fast forwarding to the present day, we observe a further significant expansion
to the range of stochastic models for which temporal logics and
model checking algorithms have been developed~\cite{KNP22}:
partially observable variants of MDPs, widely used in AI and planning,
allow modelling of autonomous systems with unreliable sensors
or of security protocols deploying secret keys;
stochastic games provide reasoning about agents operating
either competitively or collaboratively in probabilistic settings;
and uncertain (or robust) Markov models formally capture
epistemic uncertainty resulting, e.g., from data driven modelling.

To back up these advances in modelling formalisms and property specification languages,
tool support has evolved and is now readily available.
This comes in the form of both mature, general-purpose probabilistic model checkers,
such as PRISM~\cite{KNP11}, Storm~\cite{HJK+22} or the Modest toolset~\cite{HH14},
and various specialised verification tools~\cite{ABB+24}.
A multitude of larger tool-chains and software frameworks have also been built,
as probabilistic model checkers have become more robust
and offer a variety of file formats and APIs with which to interface.
The availability of this software,
combined with the breadth of available formalisms,
have seen probabilistic model checking used in a
large and diverse range of applications, both within and outside computer science.

In this paper, we provide an overview of the broad range of problems to which probabilistic model checking has been applied, with a view to showcasing the current state of the field
and giving some insight into its history.
We identify a number of key areas in which the techniques have been
particularly successful and consider how these have evolved over time,
from the early work in the field to the present day.
We also draw out the particular characteristics of the methods
that make it well suited to each application area,
and mention some of the key developments in theories and technologies that have been leveraged.
Throughout, we highlight various case studies,
illustrating the different ways in which probabilistic model checking
can be deployed, the types of insight it can bring and the diversity of its usage.
We hope that this paper therefore also serves as a guide
to both current and potential users
when selecting a formalism and analysis method for a given target application.

%% file: areas.tex
\vspace*{-0.7em}
\section{Applications of Probabilistic Model Checking}\label{sect:areas}

\begin{figure}[!t]
\begin{center}
\includegraphics[scale=0.5]{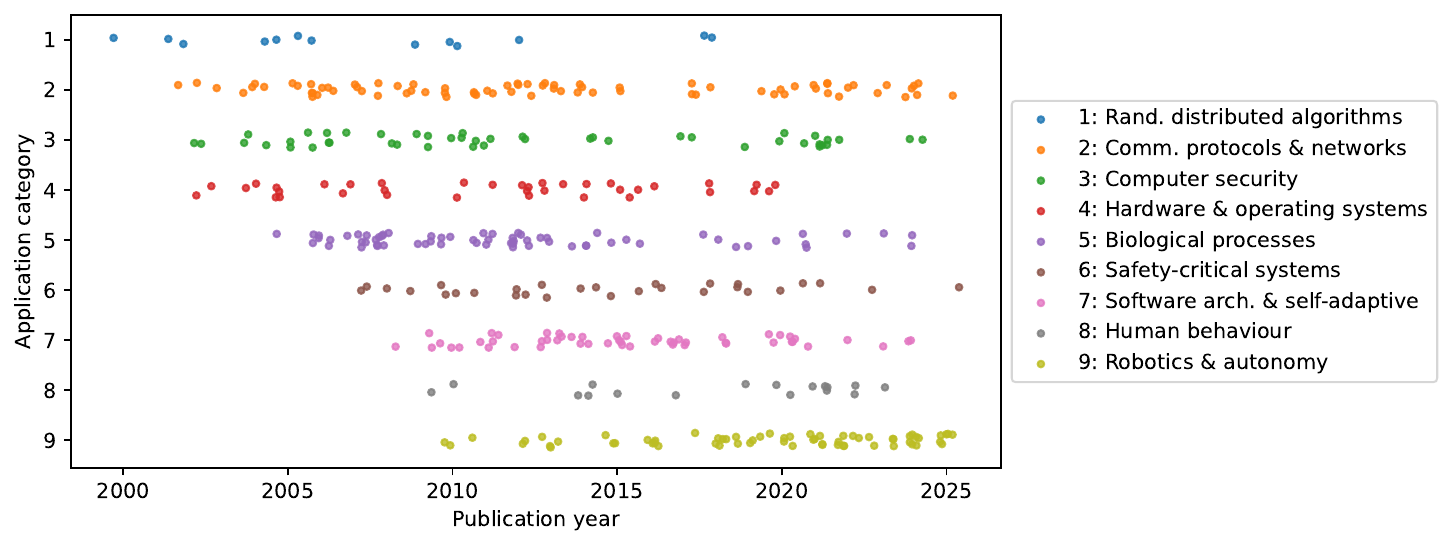}
\end{center}
\vspace*{-0.7cm}
\caption{Categorisation of approximately 400 application-oriented publications from the PRISM bibliography~\cite{prismbib}, illustrating how applications have evolved.}\label{fig:prismbib}
\end{figure}

In the following sections we discuss, in loosely chronological order,
a selection of key application areas for probabilistic model checking.
For context, Figure~\ref{fig:prismbib} presents a graphical illustration of
how these areas have evolved over time.
For this visualisation, we considered application-oriented publications from the online bibliography
of the PRISM model checker~\cite{prismbib}, in which many of the examples mentioned in
this paper can be found. We categorise them according to the areas discussed below
and show the publications that appeared in each year.

\subsection{Randomised distributed algorithms}

Motivated by the success of existing model checking techniques
in identifying bugs and unexpected behaviour in concurrent systems,
a fruitful initial area of study for probabilistic verification
was \emph{randomised distributed algorithms}.
In this setting, the use of randomisation is a crucial tool
for breaking symmetry between concurrent processes,
in order to establish correct termination or fairness between participants.
However, the subtle interaction between the parallelisation of processes
and their probabilistic behaviour makes it challenging to formally
reason about their correctness or evaluate their efficiency.

Probabilistic model checking was used to analyse
the correctness and efficiency of various randomised distributed algorithms,
including for consensus, byzantine agreement, leader election and self-stabilisation~\cite{Nor04}.
MDPs (and the closely related model of probabilistic automata)
are ideally suited to modelling the mixture of stochasticity and concurrency that arises here.
In some cases (e.g., fully synchronous settings), DTMCs are also a useful model.

The scalability of verification techniques to these applications
was enhanced significantly by the development of
\emph{symbolic} implementations of probabilistic model checking,
based on (multi-terminal) binary decision diagrams~\cite{BCHG+97,Par02}.
While the size of the state spaces of the models grows rapidly,
the high degree of regularity and the need for a mixture of both qualitative
(``consensus is eventually established with probability 1'')
and quantitative (``what is the worst-case number of algorithm rounds required for termination?'')
properties made these applications well placed to benefit from the
advantages of symbolic model checking.

In a notable example, in 2005, McIver and Morgan studied
a self-stabilization algorithm due to Herman~\cite{Her90},
establishing results on its worst-case runtime~\cite{MM05a}.
They also made a conjecture about the initial configurations of the
network that would yield the worst-case expected time to achieve stabilisation,
and validated this empirically through probabilistic model checking.
However, a proof turned out to be elusive:
a series of subsequent papers gradually established tighter bounds
until the conjecture was finally proved more than 10 years later~\cite{BGK+16}.


\subsection{Communications and networks}

\startpara{Communication protocols}
A natural evolution from early work on distributed algorithms
was to verify \emph{communication protocols}.
Again, these often deploy randomisation to break symmetry,
for example in combination with exponential back-off schemes
to prevent collisions between data transmissions;
the Ethernet protocol is a classic example.
Moreover, particularly in wireless communication settings,
it is desirable for a formal analysis of a protocol's
reliability and efficiency to also incorporate
a stochastic model of the underlying communication medium's
unreliability, e.g., message delays or losses.

As above, both DTMCs and MDPs are useful for modelling communication protocols,
with the latter in particular allowing an analysis of the \emph{worst-case}
expected performance or runtime over any possible concurrent scheduling.
Verification efforts in this area, for example of
the CSMA/CA (carrier-sense multiple access with collision avoidance), FireWire and Zigbee protocols,
also benefited from the development of \emph{probabilistic timed automata} (PTAs)~\cite{NPS13},
providing scalable modelling and verification techniques for analysing probabilistic systems
with time-outs and deadlines.
Models of randomised protocols with complex interactions but a degree of regularity
also proved to be amenable to symbolic model checking.
A good example is the verification of the Bluetooth protocol~\cite{DKNP06},
which built and analysed multiple DTMC models with more than $10^{10}$
states in order to perform a detailed analysis of the impact of differences between
variants of the protocol specification, and of the underlying model assumptions used.

Interest in this application area has continued to this day, with a wide variety of more modern wireless
protocols being verified, for example, in the context of ad-hoc or vehicular networks.
There has been particular interest in wireless sensor networks, see e.g.,~\cite{CMBF24},
where the small scale of the hardware deployed makes analysing unreliability particularly important.

\startpara{Computer and communication networks}
More broadly, probabilistic model checking has been used to formally evaluate
the effectiveness of designs for communication and computer networks.
PTA models, in particular, have been shown to be useful for
quantifying the reliability or timeliness of transmissions across a network
in the context of uncertain delays or message losses caused by individual network components.
Examples include exploring the effectiveness of publish-subscribe messaging system~\cite{HBGS07},
optimising the quality of networked automation systems for control problems~\cite{GF06c}
and comparing wireless token-passing schemes for networks in safety-critical systems~\cite{DJKG16}. 

This application domain also showcases the benefits of another important model class:
\emph{continuous-time Markov chains} (CTMCs). These provide an alternative model of stochastic systems
in which the delays between each event are represented by negative exponential distributions.
These can be very effective in modelling the timing of, for example,
the rates with which jobs arrive or are serviced in a network queue,
or with which failures occur in hardware components.

Outside formal verification, the well established fields of
\emph{performance evaluation} and \emph{performability}
use CTMCs to model and evaluate the performance and dependability
of computer and communication networks (as well as others, such as manufacturing systems).
The coming together of these field with probabilistic model checking~\cite{BHHK10}
showed that temporal logics for CTMCs like CSL and its extensions
offer a powerful means of formally specifying
a wide range of important properties across both performance and reliability
(e.g., ``the probability of a server response taking more than 1 second is at most 0.02'',
``in the long-run, the availability of the network is at least 98\%''
or ``the mean time to failure is at least 200 hours'').
Example applications include
comparing the quality-of-service of alternative approaches to traffic shaping in wireless networks~\cite{BBOM12}
and assessing the dependability of novel topologies for optical networks~\cite{SHJ17}.



\subsection{Computer security}

Computer security is another area where formal verification has long been an important tool,
due to the need for strong guarantees on the resilience of systems to attack
and the existence of unexpected weaknesses caused by subtleties in protocol designs.
Probability is again a crucial modelling tool since the use of randomisation
is widespread, e.g., for the generation of keys or session identifiers,
or to prevent buffer overflows or DNS cache poisoning attacks.

It is natural to use MDPs for the analysis of security systems,
with nondeterminism representing the potential actions of an adversary
and probability used to model randomisation or other stochastic aspects.
The work of \cite{Ste06} represents an early example
of using MDP policy synthesis within probabilistic model checking.
It generates optimal PIN block attacks, which are unexpected sequences
of interactions with an API intended to determine the value of a PIN.

This approach can be generalised by using \emph{stochastic games},
which provide formal modelling of the behaviour of multiple agents with differing objectives,
such as an attacker and defender in a secure network.
Applications of of stochastic game model checking in this area include
performing threat analysis for real-world scenarios modelling using information from technical knowledge bases~\cite{TLE22} and verifying the resilience of collective adaptive systems to attacks~\cite{GCSG15}.

Other types of models are also widely used.
For example,~\cite{ABK+10} uses CTMC model checking to study the well-known
Kaminsky DNS cache-poisoning attack, performing a detailed analysis of the efficacy
of randomisation-based fixes that have been proposed to counter the attack.
Furthermore, in many security systems, there is an inherent trade-off between security and performance,
in which case the expressivity and flexibility offered by probabilistic temporal logics is valuable.
A good example is the work in \cite{BPA+11}, which performs a detailed analysis of 
a certified e-mail message delivery protocol, 
analysing in particular the relation between error rate and transmission cost.


\subsection{Biological processes}

CTMCs have other important modelling applications beyond their usage for performance or reliability evaluation discussed above.
A notable example is for modelling the stochastic dynamics of \emph{biological processes},
from the level of cellular reactions, such as protein-protein interactions or gene expression,
up to population-level models, for example of disease spreading.

This has proved to be a popular application of probabilistic model checking
and has sparked adoption and collaboration with fields such as computational and systems biology.
Many biological systems are relatively straightforward to represent in the PRISM modelling language
and automated translations have been developed from custom languages~\cite{sbml} or process algebras~\cite{CH09}. 

As with other uses of CTMCs, the temporal logic CSL and its extensions 
are expressive enough to specify key properties of interest for biological systems
(e.g., ``what is the expected number of molecules of protein $X$ after 5 minutes?''
or ``what is the probability that more than 10\% of the population is infected within 2 weeks?'').
However, logics such as LTL have also been used to characterise
more complex temporal aspects such as oscillatory behaviour; see, e.g.,~\cite{BG10}.

From a computational perspective, verification tools implement various efficient CTMC solution methods.
However, this area, in particular, has spurred the development of \emph{statistical model checking}~\cite{YS02},
which uses discrete-event simulation to provide approximate results to verification queries
along with statistical guarantees as to their accuracy.
This class of methods works particularly well when, as here,
systems are modelled as Markov chains (i.e., without nondeterminism)
and properties of interest are often over a finite time horizon.

In contrast to many other applications of probabilistic model checking,
which feature human-engineered systems,
the mechanisms underlying biological processes are often poorly understood,
and these techniques provide a means to validate a hypothesised model.
One example is an investigation of the impact of components of the
fibroblast growth factor (FGF) signalling pathway~\cite{KH09b},
with the outputs of verification later validated against laboratory experiments.
Other examples include
an analysis of the efficacy of hyperthermia treatment~\cite{RSLG13},
which corroborated prior beliefs as to the effectiveness of combining cancer treatment strategies,
and a study of concentration-based navigation of sperm cells~\cite{KML+18},
again validating previous experimental observations. 
Similarly, probabilistic model checking can support \emph{synthetic biology},
in which biological circuits are engineered;
a good example here is DNA computing~\cite{LPC+12}.



\subsection{Safety-critical systems}

Formal verification can be particularly beneficial for \emph{safety-critical systems},
i.e., those where failures could result in death, serious injury or other significant damage.
Probabilistic model checking can be used to rigorously quantify the likelihood
with which high-risk failures or malfunctions could occur.
This can be used for the assessment of \emph{safety integrity levels} (SILs),
which specify acceptable probabilities or rates of failure,
and which underlie various technical standards
that are either recommended or mandated for safety-critical systems.

Typically, systems are modelled as DTMCs or CTMCs, with the latter being more common
if the model is being used to quantify the \emph{rate} of failure over a specified time period,
rather than the probability of a failure at any point.
Safety specifications typically amount to relatively simple PCTL or CSL queries,
although the analysis can still be expensive if probabilities are very low
or the model comprises a large number of components.
It is worth noting that probabilistic verification tools
typically feature a variety of different methods for computing probabilities,
trading off the accuracy required (which can be high in this setting) against computational cost~\cite{BHK+20}.

An early example of the use of probabilistic model checking in this area
was its integration into the \emph{failure mode and effects} analysis of an airbag system~\cite{AFG+09}.
Two contrasting designs were checked against relevant safety standards
and the precise causes of safety violations were then identified.
Connections have also been established between
probabilistic model checking and \emph{fault tree analysis}~\cite{VSKS24}
and applied for example to
reliability analysis of railway infrastructure~\cite{WVKN22} 
and safety analysis of vehicle guidance systems~\cite{GJK+17}. 

Other safety-critical applications of probabilistic model checking
include those from \emph{medical} and \emph{aerospace} domains:
verifying pacemaker designs by integrating probabilistic models of heart behaviour and quantifying the resulting probability of correct device operation~\cite{CDKM12b};
analysing the reliability, availability and maintainability of satellite systems~\cite{HMS15};
and verifying the reliability of spacecraft designs~\cite{CMS+25}. 
We also discuss below (in Section~\ref{sec:robots}) other applications
that may be safety-critical, 
such as autonomous driving and human-robot interaction.



\subsection{Hardware and operating systems}

Many of the instances of probabilistic verification surveyed above
model systems at a relatively high level of abstraction,
e.g., the interactions between participants in a protocol or components in a network.
However, formal probabilistic modelling also provides value at
the \emph{hardware} or \emph{operating system} level.

One application is to verify that hardware circuits function reliably
even in the presence of failures of individual system components.
This typically requires model checking of a DTMC
in a fashion not dissimilar to that described above for quantifying risk in safety-critical systems.
An early illustration of this was the verification of NAND multiplexing~\cite{NPKS05},
a fault-tolerant design motivated by manufacturing defects found in the field of nanotechnology.
Probabilistic model checking illustrated the impact of slight variations in component reliability
and, in doing so, identified a flaw in an earlier analysis of the design.

A much broader range of system properties than just reliability can also be studied,
including those relating to \emph{timing}, \emph{power usage} and \emph{energy consumption}.
This makes probabilistic model checking a powerful tool
to explore design spaces and investigate trade-offs between competing characteristics.
For example, more recently, work in \cite{RMT+20}
presented a detailed illustration of the integration of probabilistic model checking
into the early design process of circuit designs using reconfigurable transistors.
They computed guarantees on delay, power dissipation and energy consumption per operation,
including various properties that would be difficult to analyse with simulation.

A similar spectrum of quantitative properties can be analysed in operating systems applications.
For example, \cite{SDK+20} uses MDPs for thermal modelling of multi-core systems
and then analyses performance-reliability trade-offs.
This is done using probabilistic model checking with \emph{energy-utility quantiles}~\cite{BDK14,BDD+14},
which incorporate  conditional probabilities and ratio constraints between cost and reward measures
to investigate the interplay between multiple objectives.


\subsection{Robotics and autonomous systems}\label{sec:robots}

In recent years, a clear trend within the use of probabilistic model checking
has been a shift towards synthesising ``correct-by-construction'' controllers or policies
for a system, based on a formal specification of its desired behaviour.
Although this differs from the more classical approach of
verifying a fixed system model to determine whether it satisfies a specification,
probabilistic model checking is already well suited to this task, notably through the use of MDPs,
with temporal logic used for controller specifications.
In parallel, there has been a growing synergy between verification 
and fields of artificial intelligence such as planning and reinforcement learning, which often also use MDPs.

An application domain where these shifts have been particularly apparent is \emph{robotics and autonomous systems}.
The use of stochastic modelling is essential here
since robots frequently operate in uncertain and dynamic environments,
due to the presence of humans or other unknown obstacles,
and imperfect or unreliable sensors and actuators.
Probabilistic model checking offers a wealth of ways to formally specify desired behaviour in this setting,
and the means to provide a probabilistic guarantee on the behaviour of generated policies.
These can be critical for safety purposes,
e.g., for robots operating in the presence of humans
or high risk environments, 
or to guarantee performance and reliability
sometimes under tight resource (e.g., battery) constraints.

The use of linear temporal logic (LTL) is common here,
in order to capture more complex temporal specifications
(e.g., ``maximise the probability of inspecting all three sensors, in any order,
whilst avoiding $X$'').
This is often combined with multi-objective model checking~\cite{EKVY08},
which allows the generation of policies that trade-off between several, competing objectives
(e.g., battery life vs. mission execution time)
or analysis of the corresponding Pareto front.

Robotics provides good examples of probabilistic model checking being embedded in real-world systems.
For example, \cite{HBJ+17} describes a major \emph{long-term autonomy} project deploying
mobile robots in everyday environments such as offices and care homes. 
Multi-objective model checking on MDPs is repeatedly applied within the robots' control software
to perform tasks effectively and reliably as the map of the environment is updated over time.
In other recent work, probabilistic model checking is used within a
planning system for \emph{autonomous underwater vehicles} (AUVs) to retrieve data from sensor networks~\cite{BSK+22}.
The resulting policies are deployed in a real-world trial
and shown to outperform existing hand-designed policies.


Robotics and autonomous systems applications also highlight
further challenges arising in verification today.
One is the integration of components that deploy \emph{machine learning}.
In~\cite{CIM+24}, a probabilistic model checking framework is proposed
that incorporates deep-learning perception techniques.
The uncertainty arising from the learning-driven components
is factored into the verification of the overall system;
this is applied to controllers for mobile robots and a
driver-attentiveness management system for shared-control autonomous cars.



\subsection{Software architectures and self-adaptive systems}

The performance and reliability of \emph{software architectures}
has also been investigated quite extensively with probabilistic model checking.
In similar fashion to other applications already discussed,
such as computer and communication networks or hardware designs,
formal modelling and analysis of Markov chains
can be used to rigorously establish how the likelihood of failures or delays in individual components
impacts the overall reliability and performance of a complex system of components.
In an early illustration of this, researchers from ABB used DTMC models to evaluate
the reliability of a large-scale industrial control system with more than 100 components~\cite{KSB10}.

In recent years, there has been particular interest in the modelling
and verification of \emph{cloud computing} services,
where there is a need to maintain high Quality-of-Service levels,
even under variable and unpredictable workloads.
Moreover, it is often essential to meet precisely specified \emph{service level agreements},
relating to response time or availability, which can be conveniently expressed
and verified using probabilistic model checking; see e.g.,~\cite{KKS17}.
Perhaps more importantly, these techniques can also support
run-time decision making in cloud systems such as for auto-scaling (adjusting resource
levels to meet demand) and load balancing.
As discussed for autonomous systems above,
MDP-based model checking provides an effective way to do this,
particular when reasoning about trade-offs between different objectives or metrics; see e.g.,~\cite{NJI+24}.

More generally, \emph{self-adaptive} systems, which monitor their environment
and adapt their behaviour accordingly at run-time,
have also proved to be a popular application domain for these techniques~\cite{MCGS15}.
In fact, a variety of more advanced methods have been used effectively here.
This includes model checking of \emph{stochastic games},
which have been used to model worst-case assumptions about
the environment of the cloud system \cite{CGS015}.
Another example is the use of \emph{parametric model checking}~\cite{JAH+24},
which can in this context be used to apply a form of sensitivity analysis,
quantifying the impact that unknown system parameters have on overall performance metrics~\cite{ACPM22}.
Lastly, methods for \emph{families} of probabilistic models~\cite{CDKB18}
have been used to efficiently verify multiple configurations supported by a self-adaptive system~\cite{PBD+22}.


\subsection{Human behaviour modelling}

The modelling and analysis of \emph{human behaviour} represents another challenge for formal verification,
and one where probabilistic techniques play a key role.
An example is the evaluation of diabetes patients' behaviour when using insulin pumps~\cite{CFR+15}.
Machine learning is used to extract representative patient behavioural patterns from a clinical dataset,
and probabilistic model checking is deployed to analyse how different behavioral patterns impact an individual's glucose physiology. The results demonstrate that switching behaviour types can significantly improve a patient’s glycemic control outcomes, boosting the effectiveness of
diabetes patient education and peer support.
The work of \cite{ZBD+20} uses DTMCs and probabilistic model checking, together with a cognitive
reliability and error analysis method to transform expert estimates of relevant environmental
and cognitive factors into human error rates, to assess the reliability
of the procedures of a pharmacy and also analyse the effects of potential
alternatives.

Another interesting application is the analysis of user interactive systems~\cite{ACH+14}. The approach is based on first inferring DTMC models of users' activity patterns by applying machine learning to logged user traces. Probabilistic model checking is then applied to the DTMCs to express hypotheses about
user behaviour and relationships within and between the activity patterns. The approach has been applied to a real-world case study of a deployed app with thousands of users and the 
analysis performed revealed insights into real-life app usage.

Other applications 
to human modelling include air traffic control~\cite{RBC+13}, controller synthesis of UAVs interacting with human operators~\cite{LCHT15}, human in the loop self-adaptive systems~\cite{LAKG20}, and the design of correct-by-construction Advanced Driver Assistance Systems (ADAS) interacting with a human driver model built using the cognitive architecture ACT-R \cite{ELK19}. 


\subsection{Further application areas}

In the above, we have identified some of the most well-studied categories of applications for probabilistic model checking.
This is of course non-exhaustive, and various other applications have been considered multiple times.
These include: \emph{smart grids}, with a focus on performance, resilience or security;
\emph{quantum computing}, for example, key distribution algorithms;
\emph{blockchain}-based systems, such as bitcoin;
and \emph{business process} modelling.
See~\cite{prismbib} for details.


We conclude this discussion by highlighting a few of the more diverse applications
that have appeared in recent years, adding support for our argument that the techniques have very broad applicability.

\startpara{Sports tactics}
One example is the work in~\cite{CRD+23},
which develops a framework for reasoning about effectiveness of team strategies in professional football.
The first phase is to learn an MDP model, from event stream data,
capturing the probability of moving between areas of the pitch
and executing actions, such as passes or shots.
Then probabilistic model checking is applied, with LTL used to express outcomes of interest.
The analysis yields insights, such as when and where passing and shooting is more effective.

\startpara{Court interactions}
Another example is an analysis of interactions between participants in cases in the US Supreme Court~\cite{DS23b}.
Based on a dataset of court transcripts, a DTMC model is constructed
that models the dynamics of interactions between people over the course of a trial.
Probabilistic model checking is then used to analyse a wide range of properties,
from the expected timing of court processes, to the likelihood of various
events, such as the decisions taken by judges, and various trends are identified.

\vspace*{0.7em}
Other interesting applications include
estimating political affinities through opinion diffusion on Twitter~\cite{SGDE20},
studying the impact of regulations on the performance of public transport~\cite{BBH+19},
planning railway infrastructure through capacity evaluation~\cite{EN24} 
and assessing graphical user interfaces~\cite{BM09}.

%% file: concl.tex
\section{Conclusions and Outlook}\label{sect:concl}

We have taken a retrospective look at some of the
successful applications of probabilistic model checking over the last 25 years,
identifying commonly studied application domains
and discussing why and how they are amenable to this approach.
Commonalities exist between the different problem domains,
but there is considerable breadth and diversity to the set of applications.

Broadly speaking, we observe that this is largely because
probabilistic model checking offers ease of rigorous modelling
for many different types of stochastic behaviour,
combined with extremely flexible logical formalisms
to specify their quantitative characteristics.
These go well beyond the classical notions of correctness
typically used in formal verification.
This is backed up by a wide range of flexible analysis techniques and stable tool support.
These tools are continually evolving, incorporating new advances
in modelling and verification techniques,
which we anticipate leading to further application domains in the future. 

\vspace*{0.7em}

\startpara{Acknowledgements}
This project was funded by the ERC under the European Union’s Horizon 2020 research and innovation programme (\href{http://www.fun2model.org}{FUN2MODEL}, grant agreement No.~834115).